\documentstyle[aps,prl,multicol]{revtex}

\renewcommand{\baselinestretch}{1}
\begin{document}
\title{Quantum superposition of multiple clones and the novel cloning machine}
\author{Arun Kumar Pati}
\address{SEECS, Dean Street, University of Wales, Bangor LL 57 1UT, UK}
\date{\today}
\maketitle
\def\ra{\rangle}
\def\la{\langle}
\def\ver{\arrowvert}
\begin{abstract}
We envisage a novel quantum cloning machine, which takes an input state
and produces an output state whose success branch can exist in a linear superposition of
multiple copies of the input state and the failure branch exist in a
superposition
of composite state independent of the input state. We prove that
unknown non-orthogonal states
chosen from a set $\cal S$ can evolve into a linear superposition of multiple clones
by a unitary process if and only if the states are linearly independent. 
We derive a bound on the success
probability of the novel cloning machine. We argue that the deterministic
and probabilistic clonings are special cases of our novel cloning machine.
\end{abstract}

\vskip .5cm

PACS           NO:     03.67.-a, 03.65.Bz, 89.70.+c\\
email:akpati@sees.bangor.ac.uk\\

\vskip 1cm

\begin{multicols}{2}

\par
  The information theoretical approach to the foundations of quantum theory is
to regard
the debatable issues as the facts of the quantum world and to exploit these
facts in a constructive way so as to achieve {\em classically forbidden}
technological
applications in quantum information processing.
In recent years the quantum mechanical principles such as
linearity, unitarity and inseparability have been utilised to realise quantum
computers \cite{dd}, quantum teleportation \cite{cbb}, quantum cryptography
\cite{chb} and so on. In one hand these principles enhance the possibility of
information processing and on the other they put some limitations, too. 
That an unknown quantum state cannot be
perfectly copied is a consequence of linearity of quantum theory
\cite{wz,dk}. Later it \cite{hy} was shown that the unitarity of the quantum
theory does not allow to clone two non-orthogonal states and
thus it is impossible to measure the wavefunction of a single quantum
system \cite{ay}.
For mixed states a generalisation of  ``no-cloning'' theorem \cite{bcf}
says that there is no physical means
for broadcasting an unknown quantum state onto two separate parties.
Though perfect copies cannot be produced, there exist the 
possibility of producing approximate copies of an unknown quantum state
\cite{bhh,bbhb}.
Optimal and universal quantum cloning machines have been
constructed 
and it was shown \cite{gm} that if the device can make infinite number of copies
then it is as good as a classical copying machine.
The universal and state dependent quantum cloning machines  have
been  studied and applied to quantum cryptography \cite{be}.
Remarkably, if one allows unitary and
measurement processes then  a set of  linearly independent non-orthogonal
states can be cloned perfectly with a non-zero probability \cite{dg,dgc}.
Recently, we \cite{akp} have proposed a protocol for producing copies and
complement copies of an unknown qubit using minimal communication from a
state preparer. The problem of quantum cloning
can be regarded as a special case of state separation and unambiguous
discrimination of non-orthogonal quantum states
\cite{cb}. As some applications one finds
the possibility of decompressing quantum entanglement using local copying
\cite{bvpk}. Interestingly, the quantum ``no-cloning'' theorem  has also
been invoked to explain the information loss in side a black hole \cite{su}.

In the
past various authors have asked the question: If we have an
unknown state $\ver \psi \ra$ is there a device which
will produce either $\ver \psi \ra \rightarrow \ver \psi \ra^{\otimes 2}$,
$\ver \psi \ra \rightarrow \ver \psi \ra^{\otimes 3}$,
$\ver \psi \ra \rightarrow  \ver \psi \ra^{\otimes M}$ or
$\ver \psi \ra^{\otimes N} \rightarrow \ver \psi \ra^{\otimes M}$ copies
of an unknown state in a deterministic or
probabilistic fashion. This is a ``classicalised'' way of thinking about a
quantum cloning machine. If we pause for a second, and think the working
style of a classical Xerox machine, then we know that it does exactly the same
thing. If we feed a paper with some amount of information into a Xerox machine
containing $M$ blank papers, we can either get 
$1 \rightarrow  2$, or
$1 \rightarrow 3$, or
$1 \rightarrow  M$ copies by
just pressing the number of copies we want. However, the quantum world is
different where one can have linear superposition of all possibilities with
appropriate probabilities.
 If a real quantum cloning machine would exist it should
exploit this basic feature of the quantum world and it should produce
simultaneously $\ver \psi \ra \rightarrow \ver \psi \ra^{\otimes 2}$,
$\ver \psi \ra \rightarrow \ver \psi \ra^{\otimes 3}$, {\it and}
$\ver \psi \ra \rightarrow \ver \psi \ra^{\otimes M}$ copies. We ask if it is possible by some physical process
 to produce an output state of an unknown quantum state which will be {\it in
  a linear superposition of all possible multiple copies each
 in the same original state?} A device that can perform this task we call
 ``novel quantum cloning machine'' (NQCM).

 In this letter we show that the non-orthogonal states secretly chosen
from a set ${\cal S} = \{\ver \psi_1 \ra, \ver \psi_2 \ra,.... \ver \psi_k \}$
can evolve into a linear superposition of  multiple
copy states together with a failure term described by a composite state
 (independent of the input state) by a unitary process if and only if the states are
 linearly independent. Further, we show that the von Neumann measurement of
 a ``Xerox number'' operator can yield a distribution of perfect copies
 according to the standard rules of quantum theory.
 We also prove a bound on the success probability of the novel 
 cloning machine for non-orthogonal states.
 We point out that the
 recently proposed probabilistic cloning machine  of Duan-Guo \cite{dgc} can
 be thought of as a special case of a more general ``novel cloning machine''.
 We hope that the existence of such a machine
 would greatly facilitate the quantum information processing
 in a quantum computer.

  Consider an unknown
 input state  $\ver \psi_i \ra$ from a set $\cal S $
 which belongs
 to a Hilbert space ${\cal H}_A = C^{N_A}$. Let $\ver \Sigma \ra_B$ be
 the state of
 the ancillary system $B$ ( analogous to bunch of blank papers) which belongs to
 a Hilbert space ${\cal H}_B$ of dimension $N_B = N_A^{M}$, where $M$ is
 the total number
 of blank states each having dimension $N_A$. In fact we can take
 $\ver \Sigma \ra_B = \ver 0 \ra^{\otimes M} $. Let there be a probe state
 of the cloning device which can measure the number of copies that have been
 produces and $\ver P \ra$ be the initial state of the probing device.
 Let $\ver P_1 \ra,..., \ver P_M \ra,....\ver P_{N_C} \ra$
are orthonormal basis states of the probing device. The set $\{ \ver P_n \ra \}
\in {\cal H}_C = C^{N_C}$ such that $N_C > M $. If a novel cloning machine
exist, then it should be represented by a linear, unitary operator  that
acts on the combined
sates of the composite system.  The question is: Is it possible to have
a quantum superposition of the multiple clones of the input state given by

\begin{eqnarray}
&& \ver \psi_i \ra \ver \Sigma \ra \ver P \ra \rightarrow
U(\ver \psi_i \ra \ver \Sigma \ra \ver P \ra) = 
 \sqrt{ p_1^{(i)}} \ver \psi_i \ra \ver \psi_i \ra \ver 0 \ra ...\ver 0 \ra \ver P_1 \ra
 \nonumber \\
&& + \sqrt{ p_2^{(i)}} \ver \psi_i \ra \ver \psi_i \ra \ver \psi_i \ra ...\ver 0 \ra \ver P_2 \ra 
+.... \nonumber \\
&& + \sqrt{ p_M^{(i)}} \ver \psi_i \ra \ver \psi_i \ra ..\ver \psi_i \ra \ver P_M \ra,
\end{eqnarray}
where $p_n^{(i)}, (n=1,2,...M)$ is the probability with which $n$-copies of the
original input quantum state can be produced. However, we \cite{ap}
have recently shown
that such {\it an ideal novel cloning machine based on unitarity of
quantum theory cannot exist}. The cloning
machine should
fail some time and the failure branch should be described by a state
independent of the input state. Nevertheless, novel cloning machines which can
create linear superposition of multicopies with non-unit total
success probability do exists.
The existence of such a machine is proved by the
following theorem.

{\sl {\bf Theorem} }:~~  There exists a unitary operator $U$ such that
for any unknown state chosen from  a set ${\cal S} =
 \{\ver \psi_i \ra \}(i =1,2,..k)$ the machine  can
 create a linear superposition of multiple clones together with failure copies
 given by

\begin{eqnarray}
&& U(\ver \psi_i \ra \ver \Sigma \ra \ver P \ra) =
\sum_{n=1}^M 
\sqrt{ p_n^{(i)} } \ver \psi_i \ra^{\otimes (n+1)} \ver 0 \ra^{\otimes (M-n)} \ver P_n \ra \nonumber \\
&& + \sum_{l=M+1}^{N_C}  \sqrt f_l^{(i)} \ver \Psi_l \ra_{AB} \ver P_l \ra,
\end{eqnarray}
if and only if the states
$\ver \psi_1 \ra, \ver \psi_2 \ra,.... \ver \psi_k \ra$ are linearly independent.
In the above equation
$p_n^{(i)}$ and  $f_l^{(i)}$ are success and failure probabilities  for the $i$th
input state to produce $n$ exact copies and to remain in the $l$th failure component,
respectively.
The states $\ver \Psi_l \ra_{AB}$'s are normalised states of the composite
system $AB$ and they are not necessarily orthogonal.

 We prove the existence of such a unitary operator in two stages.
First, we prove that if
an unknown quantum state chosen from a set $\cal S$ exists in a linear superposition of
multiple copy states then the set $\cal S$ is linearly independent.
Second, (which is the converse of the above statement) we prove that if the
set ${\cal S} = \{\ver \psi_i \ra \}$ is linearly independent then any state
chosen from a set ${\cal S}$ can evolve into a linear superposition of
multiple clones.
Consider
an arbitrary state $\ver \psi \ra = \sum_{i} c_i \ver \psi_i \ra$. If we feed 
this state, then, the unitary evolution yields

\begin{eqnarray}
&& U(\ver \psi \ra \ver \Sigma \ra \ver P \ra) =
\sum_{n=1}^M 
\sqrt p_n \ver \psi \ra^{\otimes (n+1)} \ver 0 \ra^{\otimes (M-n)} \ver P_n \ra \nonumber \\
&& + \sum_{l=M+1}^{N_C}  \sqrt f_l \ver \Psi_l \ra_{AB} \ver P_l \ra.
\end{eqnarray}
However, by linearity of quantum theory each of $\ver \psi_i \ra$
would undergo transformation under (2) and we have

\begin{eqnarray}
&& U(\sum_i c_i \ver \psi_i \ra \ver \Sigma \ra \ver P \ra) =
\sum_i~c_i  \sum_{n=1}^M 
 \sqrt p_n^{(i)} \ver \psi_i \ra^{\otimes (n+1)} \ver 0 \ra^{\otimes (M-n)} \ver P_n \ra \nonumber \\
&& + \sum_i~c_i \sum_{l=M+1}^{N_C}  \sqrt f_l^{(i)} \ver \Psi_l \ra_{AB} \ver P_l \ra .
\end{eqnarray}
Since the final states in (3) and (4) are different a quantum state
represented by $\ver \psi \ra$
cannot exist in a linear superposition of all possible copy states. We know
that if a set contains distinct vectors 
$\{ \ver \psi \ra,  \ver \psi_1 \ra, \ver \psi_2 \ra,..\ver \psi_k \ra \}$ such that
 $\ver \psi \ra$ is a linear combination
  of other $\ver \psi_i \ra$'s then the set is linearly dependent. Thus
  linearity prohibits us creating linear superposition of multiple copy
  states chosen from a linearly dependent set. Therefore, the unitary (linear)
  evolution (2) exists for any state secretly
  chosen from $\cal S $ only if its elements are linearly independent. This
  proves the first part of the theorem.

    Now we prove
  the converse of the statement, i.e., we show that if the set $\cal S $ is
  linearly independent then there exists a unitary evolution, which can create
  linear superposition of multiple copy states with some success and failure.
  If the unitary evolution (2) holds, then the overlap of two distinct output
  states $\ver \psi_i \ra$ and $\ver \psi_j \ra$ secretly chosen from ${\cal S}$
  after they have passed through the device would be given by

\begin{equation}
\la \psi_i \ver \psi_j \ra  = 
\sum_{n=1}^M 
\sqrt{p_n^{(i)} }\la \psi_i \ver \psi_j \ra^{n+1} \sqrt{ p_n^{(j)} }
+ \sum_{l=M+1}^{N_C} 
\sqrt{ f_l^{(i)}  f_l^{(j)} }.
\end{equation}
Conversely, if (5)  holds, there exists a unitary operator to
satisfy (2).
In the sequel we will prove that if the set ${\cal S}$ is linearly independent
then (5) holds.
The above equation can be generically expressed as a $k \times k$ matrix equation
 
\begin{equation}
G^{(1)}  = 
\sum_{n=1}^M 
A_n G^{(n+1)} A_n^{\dagger} + \sum_l F_l,
\end{equation}
where the matrices $G^{(1)} = [ \la \psi_i \ver \psi_j \ra ]$ is the Gram matrix,
 $G^{(n+1)} =
[ \la \psi_i \ver \psi_j \ra^{(n+1)} ], A_n = A_n^{\dagger} =
 diag( \sqrt{p_n^{(1)}}, \sqrt{p_n^{(2)}},..\sqrt{p_n^{(k)}} )$ and
 $F_l = [\sqrt{ f_l^{(i)} f_l^{(j)} }]$ . Now proving the existence of
 a unitary evolution given in (2) is equivalent to showing that (6) holds
 for a positive definite matrix $A_n$.
 It can be shown that if the states $ \{ \ver \psi_i \ra \}$
 are linearly independent, then the  Gram matrix $G^{(1)}$ is a positive definite and
 its rank is equal to the dimension of the space spanned by the vectors
 $\ver \psi_i \ra$. Similarly,
 we can show that the matrix $G^{(n+1)}$ is also positive definite. Because for
 an arbitrary vector  $ \alpha  = col( c_1, c_2,....c_k)$, we can write
 $\alpha^{\dagger}~~  G^{(n+1)}~~  \alpha  = \sum_{i,j=1}^k c_i^* c_j G^{(n+1)}_{ij}
 = \la \beta \ver \beta \ra$, where $\ver \beta \ra =
  \sum_i c_i \ver \psi_i \ra^{\otimes (n+1)}$.
 Since square of the length of a vector is positive and
 cannot go to zero (if the set is linearly independent), this shows that
 $G^{(n+1)}$ is a
 positive definite matrix. Also, the matrix $A_n$ is positive definite which suggests
 $A_n G^{(n+1)} A_n^{\dagger}$ is also a positive definite matrix. Further, we
 know that sum of positive definite matrices is also a positive definite one.
 From the continuity argument for a small enough $A_n$ the matrix
 $ G^{(1)}  - \sum_{n=1}^M  A_n G^{(n+1)} A_n^{\dagger} $ is also a positive
 definite matrix. Therefore, we can
 diagonalise  the Hermitian matrix  by a suitable unitary operator $V$.
  Thus we have
  $V^{\dagger} ( G^{(1)} - \sum_n A_n G^{(n+1)} A_n^{\dagger}) V
   = diag(a_1, a_2,...a_k) $, where
 the eigenvalues $\{ a_i \}$ are positive real numbers. Now we can chose the matrix $ F_l$
 to be $F_l = V diag( g_{(l)1}, g_{(l)2},....g_{(l)k} ) V^{\dagger}$ such that
 $\sum_l g_{(l)i} = a_i, (i =1,2,...k) $. Thus
 the matrix equation (5)
 is satisfied with a positive definite matrix $A_n$ if the states are linearly independent.
 Once (2) holds, we see
 that the success and failure probabilities are summed to unity,
 i.e. $\sum_n p_n^{(i)} + \sum_l f_l^{(i)} = 1$ as expected.
 This completes the proof of our main result.

   Here we discuss the generality of our novel cloning machine.
   For example, if  $\ver 0 \ra$ and $\ver 1 \ra$
  are the computational basis, then a qubit secretly chosen from a
 set $\{ \ver 0 \ra, \alpha \ver 0 \ra
 + \beta \ver 1 \ra \}$ or from a set $\{ \ver 1 \ra, \alpha \ver 0 \ra 
 + \beta \ver 1 \ra \}$ 
 can exist in a linear superposition of multiple clones. But a state chosen from
 a  set $\{ \ver 0 \ra, \ver 1 \ra, \alpha \ver 0 \ra  + \beta \ver 1 \ra \}$ 
 cannot exist in such a superposition of multiple clones as the set is not linearly
 independent. It may be remarked that ``no-cloning'' theorem is a special case
 of our result. Because, the linear superposition of multiple
 clones fails if the machine does not fail with some probability. When all the
 failure probabilities are zero, we have ``no-superposition of
 multi-clones'' theorem \cite{ap}. Then if we take one of the success probability as
 one, then we get  Wootters-Zurek-Diek's ``no-cloning'' theorem \cite{wz,dk}.

  Our result is consistent with the known results on cloning. In the unitary
  evolution if one of the positive number in success branch is one (i.e.
  $p_n^{(i)} = 1$ for some $n$ and all $i$) and all
  others (including failure branches) are zero, then
  we have
  $U(\ver \psi_i \ra \ver \Sigma \ra \ver P \ra) =
  \ver \psi_i \ra^{\otimes (n+1)} \ver 0 \ra^{\otimes (M-n)} \ver P_n \ra$.
   This tells us that
  the matrix equation would be $G^{(1)} = G^{(n+1)}$ since $A_n = I$. This
  will be possible only when the states chosen from a set are orthogonal
 to each other. Thus a single quanta
 in an orthogonal state can be perfectly cloned \cite{hy}.
 Here we discuss the condition under which all $f^{(i)}$'s are zero.
 The orthogonality relation
 $\la \psi_i \ver \psi_j \ra = \delta_{ij}$
 is a necesary and sufficient condition on the set ${\cal S}$ for all
 $f^{(i)}$'s to be zero. The converse
 can be also proved, i.e., if all $f^{(i)}$'s are zero then the states are
 orthogonal. When all $f^{(i)}$'s are zero, from (5) we can obtain
 $\ver \la \psi_i \ver \psi_j \ra \ver \le
 \sum_{n=1}^M  \sqrt{ p_n^{(i)} p_n^{(j)} } \ver \la \psi_i \ver \psi_j \ra \ver^{n+1}$.
 Using two inequalities $ (p_n^{(i)} . p_n^{(j)} )^{1 \over 2} \le {1 \over 2}
(p_n^{(i)} + p_n^{(j)} )$  and $\ver \la \psi_i \ver \psi_j \ra \ver^{n}
  \le  \ver \la \psi_i \ver \psi_j \ra \ver$ (the later is valid for any two
 states) we find $\ver \la \psi_i \ver \psi_j \ra \ver (1- 
 \ver \la \psi_i \ver \psi_j \ra \ver) \le 0 $. Since the quantity in side
 the bracket is positive and $\ver \la \psi_i \ver \psi_j \ra \ver$ cannot
 be negative it must be zero. Therefore, the
 states have to be orthogonal when all $f^{(i)}$'s  are zero.
 Note that another interesting result follows from our proposed cloning machine.
 If the states are orthogonal and all $p_n^{(i)}$'s are non-zero, then unitarity
 allows us to have a {\it linear superposition of multiple copies of
 orthogonal states } as the matrix
 equation is always satisfied. We mention that it would be interesting to
 investigate the extension of $U$ beyond the elements of ${\cal S}$ in future.

 After the input state chosen from the set $\cal S$ undergo unitary evolution
 in order to know how many copies are produced by the novel cloning machine,
  one needs to do a von Neumann
 measurement onto the probe basis. This can be thought of as a measurement
 of a Hermitian operator. We introduce such an operator, which is called
 ``Xerox number'' operator $N_{X}$, defined as

\begin{equation}
N_X = \sum_{n=1}^{M}~~n \ver P_n \ra \la P_n \ver.
\end{equation}
The probe states $\ver P_n \ra$ are eigenstates of the Xerox number operator
with eigenvalue $n$  where $n$ is the number of clones produced with a
probability distribution $p_n^{(i)}$.
The measurement of Xerox number operator will give us information about how
many copies have been produced by the cloning machine.
For example, the novel cloning machine
would produce $1 \rightarrow 2$ copies with probability $p_1$,
$1 \rightarrow 3$ copies with probability $p_2$,
...and $1 \rightarrow M+1$ copies with probability $p_M$ in accordance
with the usual rules of quantum mechanics.

   Here, we derive a bound on the success probability of producing
multiple clones through a unitary machine (2). Taking the obverlap of
two distinct states we find

\begin{equation}
\ver \la \psi_i \ver \psi_j \ra \ver
\le
\sum_{n=1}^M 
\sqrt{ p_n^{(i)} p_n^{(j)} } \ver \la \psi_i \ver \psi_j \ra \ver^{n+1}
+
\sum_{l=M+1}^{N_C}
\sqrt{ f_l^{(i)}  f_l^{(j)} } .
\end{equation}

On simplifying (8) we get
the tight bound on the individual success
probability for cloning of two distinct non-orthogonal states as

\begin{equation}
{1 \over 2} \sum_n (p_n^{(i)} + p_n^{(j)} )
(1 - \ver \la \psi_i \ver \psi_j \ra \ver^{n+1} ) \le
(1 - \ver \la \psi_i \ver \psi_j \ra \ver ).
\end{equation}

The above bound  is related to the distinguishable
metric of the quantum state space. Since
the ``minimum-normed-distance'' \cite {ak} between two non-orthogonal states  
$\ver \psi_i \ra$ and $ \ver \psi_j \ra $ is $D^2( \ver \psi_i \ra,
 \ver \psi_j \ra) = 2 (1 -
 \ver \la \psi_i \ver \psi_j \ra \ver )$ and the ``minimum-normed-distance''
between $n+1$ clones  is
$D^2(\ver \psi_i \ra^{\otimes n+1}, \ver \psi_j \ra^{\otimes n+1}) = 2 (1 - 
\ver \la \psi_i \ver \psi_j \ra \ver^{n+1} )$, the tight bound can be
expressed as

\begin{equation}
\sum_n p_n D^2(\ver \psi_i \ra^{\otimes n+1}, \ver \psi_j \ra^{\otimes n+1})
\le D^2(\ver \psi_i \ra, \ver \psi_j \ra),
\end{equation}
where we have defined total success probability $p_n$ for $n$th clones
as $p_n = {1 \over 2} (p_n^{(i)}+  p_n^{(j)})$. The ``minimum-normed-distance''
function is a measure of distinguishability of two non-orthogonal quantum
states.
Therefore, the above bound can be interpreted physically as the
sum of the weighted distance between
two distinct states of $n+1$ clones is always bounded by the the original
distance between two non-orthogonal states.
Since cloning transformation is a physical
procedure for making two states more distinguishable, any two state which
pass through our machine has to satisfy this strict inequality. Also, our
bound is consistent with the known results on cloning. For example, if
we have $ 1 \rightarrow 2$ cloning, then in the evolution we have 
 $p_1^{(i)}$ and
$p_1^{(j)}$  are non-zero
and all others are zero. In this case our bound reduces to
$ {1 \over 2} (p_1^{(i)} + p_1^{(j)} )
\le
{1 \over 1 + \ver \la \psi_i \ver \psi_j \ra \ver }$,
which is nothing but the Duan-Guo bound \cite{dgc} for producing two clones in a
probabilistic fashion. Similarly if we have 
$ 1 \rightarrow M$ cloning, then in the evolution we have 
 $p_M^{(i)}$ and
$p_M^{(j)}$  are non-zero
and all others are zero. In this case our bound reduces to
${1 \over 2} (p_M^{(i)} + p_M^{(j)} )
\le
{1 - \ver \la \psi_i \ver \psi_j \ra \ver
\over 1 -  \ver \la \psi_i \ver \psi_j \ra \ver^M }$,
which is nothing but Chefles-Barnett \cite{cb} bound, obtained using quantum
state separation method.

   We can imagine a more general novel cloning machine and
then show that the probabilistic cloning machine discussed by Duan and Guo \cite{dgc}
can be considered as a special case of the general novel cloning machine.
Instead of the unitary evolution (2) one could describe a general unitary
evolution of the composite system $ABC$ as

\begin{eqnarray}
&& U(\ver \psi_i \ra \ver \Sigma \ra \ver P \ra) =
\sum_{n=1}^M 
\sqrt{ p_n^{(i)} } \ver \psi_i \ra^{\otimes (n+1)} \ver 0 \ra^{\otimes (M-n)} \ver P_n \ra \nonumber \\
&& + \sum_l c_{il} \ver \Psi_l \ra_{ABC}.
\end{eqnarray}

Here, the first term has the usual meaning and the second term represents the
failure term. The states $\{ \ver \Psi_l \ra \} \in {\cal H}_A \otimes
{\cal H}_B \otimes {\cal H}_C $ are normalised states of the composite system.
For simplicity we assume that they are orthonormal. Further, 
since the measurement of Xerox number operator should yield perfect copies
(say n) of the input state with probability $p_n^{(i)}$, this entails that
$\ver P_n \ra \la P_n \ver \Psi_l \ra_{ABC} = 0$ for any $n$ and $l$. Imposing
this physical condition, we find from eq.(11) that the inner product of two
distinct states gives
$\la \psi_i \ver \psi_j \ra  = 
\sum_{n=1}^M 
\sqrt{p_n^{(i)} }\la \psi_i \ver \psi_j \ra^{n+1} \sqrt{ p_n^{(j)} }
+ \sum_{l} c_{il}^{*} c_{jl}$
This can be expressed as a matrix equation
$ G^{(1)}  = \sum_{n=1}^M  A_n G^{(n+1)} A_n^{\dagger} + C^{\dagger}C$,
where $C =[c_{ij} ]$. From our earlier theorem we can now prove that
with a positive definite matrix $A_n$ we can diagonalise 
$G^{(1)}  - \sum_{n=1}^M A_n G^{(n+1)} A_n^{\dagger}$ and with a
particular choice of
the matrix $C$ the unitary evolution exists.

      To see that from our machine Duan-Guo machine follows as a special
case, let us take one of the $p_n^{(i)}$ is
non-zero and all others are zero in the above unitary transformation.
Then we have the following evolution for
the non-orthogonal states

\begin{equation}
U(\ver \psi_i \ra \ver \Sigma \ra \ver P \ra) =
\sqrt{ p_n^{(i)} } \ver \psi_i \ra^{\otimes (n+1)} \ver P_n \ra
+ \sum_l c_{il} \ver \Psi_l \ra_{ABC}.
\end{equation}
where we have assumed that there are $n$ blank states. This is nothing but
Duan-Guo type of probabilistic machine for producing $1 \rightarrow n$ copies.
If one does a measurement of the probe with a postselection of the measurement
results, then this will yield $n (n=1,2,..M)$ exact copies of the unknown quantum states.
Since all other deterministic cloning machines
are special cases of Duan-Guo machine, we can say, in fact, that all deterministic
and probabilistic cloning machines are special cases of our novel
cloning machines.

     To conclude this letter, we discovered yet another surprising feature
of cloning transformation, which says that unitarity allows us to have linear
superposition of multiple clones 
of non-orthogonal states 
along with a
failure term if and only if the states are linearly independent.
We derived a tight bound on the success probability of passing
two non-orthogonal
states through a novel cloning machine.
We proved
that the probabilistic and deterministic clonings are special cases of
our novel cloning machine.
We hope that the existence of linear superposition of multiple
clones will be very much useful in quantum state engineering,
easy preservation of important quantum information, quantum error
correction and
parallel storage of information in a quantum computer. Since multiple clones
remains in various branches of the output state in a quantum computer
one might think of manipulating different
clones in a desired  way using controlled operations.

\vskip .5cm

I thank S. L. Braunstein for useful discussions. 
I thank L. M. Duan for useful discussions and suggestions.
I gratefully acknowledge financial support by EPSRC.

\renewcommand{\baselinestretch}{1}

\end{multicols}

\end{document}